\documentstyle[amscd,amssymb,verbatim,epsf]{amsart}

\begin{document}
\theoremstyle{plain}
\newtheorem{Thm}{Theorem}
\newtheorem{Cor}{Corollary}
\newtheorem{Main}{Main Theorem}
\newtheorem{Note}{Note}
\newtheorem{Lem}{Lemma}
\newtheorem{Prop}{Proposition}

\theoremstyle{definition}
\newtheorem{Def}{Definition}

\theoremstyle{remark}
\newtheorem{notation}{Notation}
\renewcommand{\thenotation}{}

\errorcontextlines=0
\numberwithin{equation}{section}
\renewcommand{\rm}{\normalshape}%

\title[Stationary Perfect Fluids]%
   {A Structure Theorem for Stationary \\ Perfect Fluids}
\author{Brendan Guilfoyle}
\address{Brendan Guilfoyle\\
          Department of Mathematics and Computing \\
          Institute of Technology, Tralee \\
          Clash \\
          Tralee  \\
          Co. Kerry \\
          Ireland.}
\email{brendan.guilfoyle@@ittralee.ie}

\date{February 21st, 2005}

\begin{abstract}
It is proven that, under mild physical assumptions, an isolated
stationary relativistic perfect fluid consists of a finite number of
cells fibred by invariant annuli or invariant tori. For axially
symmetric circular flows it is shown that the fluid consists of cells
fibred by rigidly rotating annuli or tori.
\end{abstract}

\maketitle

\section{Introduction}

The purpose of this paper is to extend some results in the
non-relativistic theory of Euler flows to the theory of 
stationary perfect fluids in general relativity. The latter play an
important role in understanding the Einstein field equations in the
presence of matter, while the former sits
within the larger field of topological hydrodynamics. In particular,
our main result is:

\vspace{0.1in}

\noindent{\bf Theorem 1}:
{\it
Consider an analytic stationary perfect fluid flow on an analytic spacetime
($^4M$,$^4g$) with proper pressure $p$ and density $\mu$ satisfying
Einstein's field equations, which has the following attributes: 

\begin{enumerate}
\item[(a)] it is spatially compact,
\item[(b)] it satisfies the weak energy condition: $p+\mu>0$,
\item[(c)] it is isentropic: $\mu=\mu(p)$,
\item[(d)] it is non-Beltrami (see Definition \ref{d:belt} below).
\end{enumerate}

Then the spatial fluid region is divided into a finite number of cells
of two types: if the cell does not intersect the boundary it is fibred by
tori invariant under the flow, if the cell intersects the boundary it
is fibred by annuli invariant under the flow. The flow lines on a
torus are either all closed or all dense, while the flow lines on an
annulus are all closed.
}

\setlength{\epsfxsize}{3.5in}
\begin{center}
   \mbox{\epsfbox{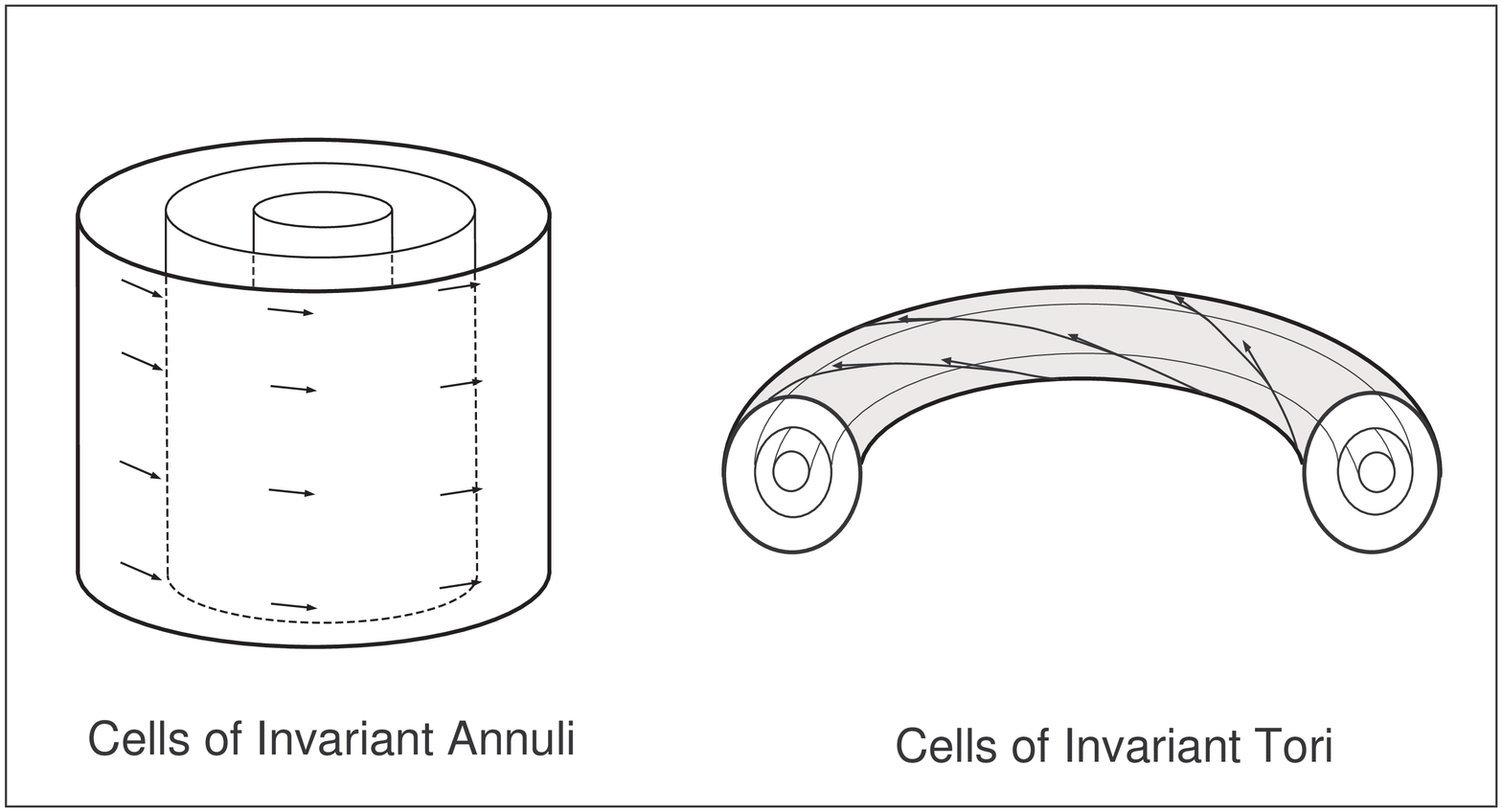}}
\end{center}

\vspace{0.1in}

To prove this theorem we find a pair of
commuting vector fields that generate the invariant surfaces.
These invariant surfaces have less geometric significance than the
surfaces of transitivity of axial symmetry, but the theorem states
that they are (generically) topologically restricted. No more than the
conservation equations of general relativity are needed to prove this. 

We apply this theorem in the case where the fluid is axially
symmetric and the fluid 4-velocity lies in the plane spanned by two
Killing vectors, so called {\it circular flows}: 

\vspace{0.1in}

\noindent{\bf Theorem 2}:
{\it
A circular fluid flow is made up of cells foliated by rigidly rotating
annuli or tori. 
}

\vspace{0.1in}

In the next section we summarise the background material of
stationary perfect fluids needed to prove the Structure Theorem. This
includes a reformulation of the conservation equations for a
relativistic  stationary perfect fluid along the lines
of the Euler equations for a steady incompressible fluid flow.
Section 3 provides the definition of relativistic Beltrami flows and
the proof of Theorem 1 while section 4 applies the theorem to
circular fluid flows. The final section contains a discussion of the
results and their implications.

\section{Stationary Perfect Fluid Flows}
A lorentz 4-manifold ($^4M, ^4g$) is said to be {\it stationary} if it
admits a timelike Killing vector $\xi^\nu$.   We can choose coordinates
($x^0=t,x^i$) on a stationary spacetime  so that $\xi^\mu=\delta^\mu_0$
and the metric takes the $t$-independent form
\[
ds^2=g_{ij}dx^i\;dx^j-e^\omega(dt+A_idx^i)^2. 
\]
Here and throughout Latin indices run through $1,2,3$, while Greek
indices run through $0,1,2,3$.

Locally we can quotient out by the action of the Killing vector on
$^4M$ to get a riemannian 3-manifold ($M,g$), which we think of as the
spatial slice of the spacetime.  We use this riemannian metric to
raise and lower all spatial tensor indices.  However, for tensors
defined on the whole spacetime we will continue to use $^4g$.  For
example, since $U^\mu$ is a 4-vector
\[
U_i=\;^4g_{i\nu}U^\nu=\;^4g_{ij}U^j+\;^4g_{i0}U^0= g_{ij}U^j-e^\omega A_i\;A_jU^j+\;^4g_{i0}U^0,
\]
while for a 3-vector $V^i$, $V_i=g_{ij}V^j$.

For later use, we note that for any $t$-independent 4-vector $U^\mu$ the divergence with respect to the lorentz and riemannian metrics are related by the simple formula
\[
^4\nabla_\mu U^\mu=\nabla_iU^i +{\frac{1}{2}} \omega_{,i}U^i.
\]
For any 4-vector  $U^\mu$ we note the following useful relations
\[
U^0=-e^{-\omega}U_0-A_iU^i \qquad\qquad U_0=-e^{\omega}(U^0+A_iU^i)
\]
\[
U^i=g^{ij}U_j-A^iU_0 \qquad\qquad U_i=g_{ij}U^j+A_iU_0,
\]
and if $U_\mu U^\mu=-1$ we have that
\[
e^{-\omega}U_0^{\;2}=1+g_{ij}U^iU^j.
\]

Consider a stationary spacetime ($^4M,$ $^4g$) on which we have a perfect fluid with 4-velocity $U^\nu$, pressure $p$, density $\mu$ and energy momentum tensor
\[
T_{\alpha\beta}=(p+\mu)U_\alpha U_\beta+p\;^4g_{\alpha\beta},
\]
and $U_\nu U^\nu=-1$.

The assumption that the flow is {\it steady} on a stationary spacetime means that  $U^\nu$, $p$ and $\mu$ are all independent of $t$.

For such a fluid, the conservation equations $^4\nabla^\mu T_{\nu\mu}=0$ read
\begin{equation}\label{e:euler1a}
^4\nabla_\nu p+2(p+\mu)U^\alpha\;^4\nabla_{[\alpha}U_{\nu]}      
      +\left(U^\alpha\;^4\nabla_\alpha(p+\mu)+(p+\mu)\;^4\nabla_\alpha U^\alpha\right)U_\nu=0.
\end{equation}
Multiplying this across by $U^\nu$ we see that
\begin{equation}\label{e:euler1aa}
(p+\mu)\;^4\nabla_\alpha U^\alpha+U^\alpha\;^4\nabla_\alpha\mu=0.
\end{equation}
Substituting this back into (\ref{e:euler1a}) we get that
\[
^4\nabla_\nu p+2(p+\mu)U^\alpha\;^4\nabla_{[\alpha}U_{\nu]}      
      +U_\nu U^\alpha\;^4\nabla_\alpha p=0.
\]
Now taking the $\nu=0$ component we get
\begin{equation}\label{e:euler1b}
U_0U^i\nabla_ip+(p+\mu)U^i\nabla_iU_0=0,
\end{equation}
while taking the $\nu=i$ component we get
\[
\nabla_ip+2(p+\mu)U^j\nabla_{[j}U_{i]}-(p+\mu)U^0\nabla_iU_0+U_iU^j\nabla_jp=0.
\]
Multiplying this across by $U_0$, using (\ref{e:euler1b}) and the fact that $U_0U^0=-1-U_iU^i$ we find that
\begin{equation}\label{e:euler2b}
U_0\nabla_ip+(p+\mu)\nabla_iU_0 +2(p+\mu)U_0^2U^j\nabla_{[j}(U_0^{-1}U_{i]}) =0.
\end{equation}
Equations (\ref{e:euler1aa}) and (\ref{e:euler2b}) are the
relativistic Euler equations governing a steady fluid perfect fluid
flow on a stationary spacetime. 

\begin{Def}
A perfect fluid is {\bf isentropic} if it has an equation of state
$\mu=\mu(p)$.   
\end{Def}

In what follows we will assume that the perfect fluid is
isentropic. If we introduce the real-valued functions $\alpha$ and
$\beta$ on $M$ given by
\[
\ln(\alpha )=\int\frac{dp}{p+\mu} \qquad\qquad \ln(\beta)=\int\frac{d\mu}{p+\mu},
\]
then the conservation equations can be written
\begin{equation}\label{e:euler1c}
\nabla_iU^i+{\frac{1}{2}}U^i\nabla_i\omega+U^i\nabla_i\ln( \beta)=0
\end{equation}
\begin{equation}\label{e:euler2c}
\nabla_i\ln[\alpha U_0]+2U_0U^j\nabla_{[j}\left(U_0^{-1}U_{i]}\right)=0.
\end{equation}

\begin{Def} An {\bf isolated} stationary body is one for which the
spatial set $V\subset M$ on which the energy momentum tensor is
non-zero is compact, and the Darmois conditions
hold  on the boundary \cite{mas1}. 
\end{Def}

For a perfect fluid (with $p+\mu\neq 0$) being isolated implies the
vanishing of the pressure and tangency of the fluid flow at the boundary.

\section{The Structure Theorem}

Consider the two vector fields
\[
X^i=U_0U^i \qquad \qquad 
Y^i=U_0\beta^{-1}e^{-\frac{1}{2}\omega}\eta^{ijk}\nabla_{[j}\left(U_0^{-1}U_{k]}\right),
\]
where $\eta$ is the hodge star operator:
\[
\eta^{ijk}=\frac{1}{\sqrt{det \;g}}\epsilon^{ijk},
\]
$\epsilon^{ijk}$ being the 3-dimensional Levi-Civita permutation symbol.

\begin{Def}\label{d:belt}
We say that the flow is {\bf Beltrami} if $Y^i=\lambda X^i$ for some
real-valued function $\lambda$ on $V$.  
\end{Def}

The reason for such a definition becomes clear from the following
structure theorem for relativistic isentropic flows:

\begin{Thm}

Consider a stationary analytic perfect fluid flow on an analytic
spacetime that: 

\begin{enumerate}
\item[(a)] is isolated,
\item[(b)] satisfies the weak energy condition $p+\mu>0$,
\item[(c)] is isentropic: $\mu=\mu(p)$,
\item[(d)] is non-Beltrami.
\end{enumerate}

Then the spatial fluid region is divided into a
finite number of cells of two types:  if the cell does not intersect
the boundary it is fibred by tori invariant under the flow, if the
cell intersects the boundary it is fibred by annuli invariant under
the flow. The flow lines on a torus are either all closed or all
dense, while the flow lines on an annulus are all closed. 

\end{Thm}

\begin{pf}

The key to the proof is to show that the vector fields $X^i$ and $Y^i$
commute. To show this we use the 3-dimensional vector notation
\[
\vec{U}\times \vec{V}=
\eta^{ijk}U_{[j}V_{k]}\frac{\partial}{\partial x^i} \qquad\qquad
\mbox{curl}(\vec{U}) =
\eta^{ijk}\nabla_{[j}U_{k]}\frac{\partial}{\partial x^i}
\]
\[
\mbox{div}(\vec{U})=\nabla_iU^i \qquad\qquad
\mbox{grad}(p)=g^{ij}\frac{\partial p}{\partial x^i}\frac{\partial}{\partial x^j}.
\]
The Euler equations (\ref{e:euler1c}) and (\ref{e:euler2c}) say that
\[
\mbox{div}(\vec{X})=\vec{X}\cdot \mbox{grad}[-{\frac{1}{2}}\omega+\ln(U_0)-\ln(\beta)]
\]
\[
\mbox{grad}\left(\ln[\alpha U_0]\right)-\vec{X}\times \mbox{curl}(\vec{T})=0,
\]
where we have introduced the 3-vectors $W^i=U^i$ and $T^i=g^{ij}U_0^{-1}U_j=U_0^{-1}W^i+A^i$.  To compute the commutator of $X^i$ and $Y^i$ we utilise the vector identity
\[
[\vec{X},\vec{Y}]=\mbox{curl}(\vec{Y}\times \vec{X})+\vec{X}\mbox{div}(\vec{Y})-\vec{Y}\mbox{div}(\vec{X}).
\]
With the help of the Euler equations and the vector identities
$\mbox{curl}(\mbox{grad} \;f)=0$ and $\mbox{div}(\mbox{curl}(\vec{T}))=0$, this says that
\begin{align}
[\vec{X},\mbox{curl}(\vec{T})]&=\mbox{curl}(\mbox{curl}(\vec{T})\times \vec{X})+\vec{X}\mbox{div}(\mbox{curl}(\vec{T}))-\mbox{curl}(\vec{T})\mbox{div}(\vec{X})\nonumber\\
& = -\mbox{curl}(\vec{T}) \vec{X}\cdot \mbox{grad}[-{\frac{1}{2}}\omega +\ln(U_0)-\ln(\beta)].\nonumber
\end{align}
This can be rearranged to
\[
[\vec{X},\vec{Y}]=[\vec{X},U_0\beta^{-1}e^{-{\frac{1}{2}}\omega}\mbox{curl}(\vec{T})]=0,
\]
as required.

The rest of the proof now follows that of the non-relativistic
case and therefore we only sketch it below - details can be found in
\cite{arnakhesin}. 

By analyticity, the critical set $K$ of the function $\ln(\alpha U_0)$
forms an semianalytic subset and $M\backslash K$ has a finite number
of cells. 

Assume that $X^i$ and $Y^i$ are not everywhere collinear.  We can
conclude that at non-critical values for $\ln(\alpha U_0) $, the level
sets form an oriented smooth 2-manifold with an ${\Bbb{R}}^2$ action.
If the level set does not intersect  $\partial V$, it forms a torus
with all flow lines closed or all flow lines dense.  If it does
intersect the boundary, it must form an annulus with closed
flowlines.

\end{pf}

{\bf Note}: A similar Structure Theorem holds in the case of dust,
where the pressure is identically zero.

\section{Axially Symmetric Flows}

When one considers the definition of axial symmetry for a stationary
field, there are a number of reasonable possibilities \cite{barnes1}
\cite{barnes2} \cite{mas2}. For our purposes we use the following
definition: 

\begin{Def}
A stationary space-time ($^4M$,$^4g$), with timelike Killing vector
$\xi^\nu$, is {\it axially symmetric} if there exists a spacelike
Killing vector $\eta^\nu$ that has closed orbits and commutes with
$\xi^\nu$.   
\end{Def}

Now suppose we have a stationary, axisymmetric spacetime ($^4M$,$^4g$)
as defined above. The simplest type of flow appears to be one where
the spatial fluid flow $U^i$ is parallel to $\eta^i$. 

\begin{Def}
A fluid flow on a stationary axisymmetric spacetime is {\it circular}
if the fluid 4-velocity $U^\nu$ lies in the plane spanned by the
Killing vectors $\xi^\nu$ and $\eta^\nu$. In addition we assume that
the spacetime has a regular axis of symmetry  \cite{carter2}
\cite{kunatru}. 
\end{Def}

\begin{Prop}
A circular flow is Beltrami iff $\lambda=0$.

\end{Prop}
\begin{pf}

Working locally there exist coordinates ($x^a,\phi,t$) such that
the Killing vectors are
\[
\vec{\xi}=\frac{\partial}{\partial t} \qquad\qquad \vec{\eta}=\frac{\partial}{\partial \phi}
\]
We now utilise a remarkable theorem \cite{carter2} \cite{kunatru} that
says that for a circular fluid flow, the 2-planes orthogonal to the
Killing orbits generated by $\xi^\nu$ and $\eta^\nu$ are
integrable. Thus the metric can be reduce to
\begin{equation}\label{e:circmet}
ds^2=h_{ab}dx^adx^b+H^2d\phi^2-e^{\omega}(dt+A\:d\phi)^2,
\end{equation}
where all of the components of the metric are independent of both $t$
and $\phi$.
 
Circularity means that the fluid flow vector is given by
\begin{equation}\label{e:Om}
U^\nu=U^0\left(\xi^\nu+\Omega\eta^\nu\right),
\end{equation}
for some function $\Omega$.  In particular, the only non-vanishing
component of the spatial vector $\vec{X}$ is 
\[
X^\phi=U_0U^0\Omega.
\]
On the other hand 
\[
Y^\phi=U_0\beta^{-1}e^{-\frac{1}{2}\omega}(det(g))^{-\frac{1}{2}}
\epsilon^{bc\phi}\partial_{[b}\left(U_0^{-1}W_{c]}+A_{c]}\right)=0.
\]
This means that $\vec{X}$ and $\vec{Y}$ are orthogonal, and so the
flow is Beltrami iff $\vec{Y}=0$. 
\end{pf}

Virtually all results on stationary axially symmetric perfect fluids,
whether exact solutions, asymptotic behaviour or {\it a prior\'{i}}
estimates, pertain only to circular flows. Thus the metric has the
form (\ref{e:circmet}) and the fluid velocity (\ref{e:Om}).

\begin{Def}
The function $\Omega$ defined in equation (\ref{e:Om}) is called the
{\it angular velocity} of the fluid.  A spinning fluid body with
constant angular velocity is said to be in {\it rigid rotation}. If
the angular velocity is not constant, the fluid is said to have {\it
differential rotation}. 
\end{Def}

\begin{Thm}
A circular flow consists of cells foliated by rigidly rotating annuli
or tori.
\end{Thm}

\begin{pf}
First, we gather together some of the
relationships enjoyed by the angular velocity, the spatial fluid
velocity vector $W^i$, the 4-components of the fluid $U^0$ and $U_0$,
and the metric functions $H$, $e^\omega$ and $A$.
\[
U_0=-(1+A\Omega)e^\omega U^0=\sqrt{1+|W|^2}\:e^{\frac{1}{2}\omega}
\]
\[
W^\phi=U^0\Omega \qquad W_\phi=\frac{|W|^2}{U^0\Omega}
\]
\[
H^2=\frac{|W|^2}{(U^0\Omega)^2} \qquad\qquad
U^0U_0=-\frac{1+|W|^2}{1+A\Omega}
\]
\begin{equation}\label{e:id5}
A=-\left(\frac{1}{\Omega}+\frac{1+|W|^2}{U^0U_0\Omega}\right),
\end{equation}
where we have introduced $|W|^2=g_{ij}W^iW^j$.

Now, introduce coordinates ($\rho$,$z$) on the 2-space
orthogonal to the group orbits, so that the commuting vector fields are
\[
\vec{X}=X^\phi\frac{\partial}{\partial \phi} \qquad
\vec{Y}=Y^z\frac{\partial}{\partial z}.
\]
Here $\rho$ labels the invariant surfaces and ($z$,$\phi$)
are coordinates on the invariant surfaces, analogous to the {\it
action angles} of classical mechanics. In these coordinates the
Euler equations dramatically simplify to 
\[
\frac{\partial}{\partial z}\left(U^0U_0\Omega\right)=0 \qquad\qquad
\frac{\partial}{\partial z}\left(A+\frac{|W|^2}{U^0U_0\Omega}\right)=0
\]
\[
\frac{\partial}{\partial z}\left(\alpha U_0\right)=0\qquad\qquad
\frac{\partial}{\partial \rho}\left(\alpha U_0\right)=
U^0U_0\Omega\frac{\partial}{\partial \rho}\left(A+\frac{|W|^2}{U^0U_0\Omega}\right).
\]
The first two of these, along with the identity (\ref{e:id5}), tell us
that the angular velocity $\Omega$ is constant on each invariant
surface. Thus each annulus or torus rotates rigidly, although there
may be differential rotation between the surfaces.  In addition,
$U^0U_0$ is constant on each invariant surface.
\end{pf}

\section{Discussion}

In topological (non-relativistic) hydrodynamics there is an equivalent
Structure Theorem for steady incompressible perfect fluid flows on a
riemannian 3-manifold \cite{arnakhesin}.  The dynamics of the fluid is
described by four equations: the three Euler equations and the
incompressibility condition. In this setting a fluid is {\it Beltrami}
if the velocity vector is everywhere collinear with it's curl. In
contrast to the regular behaviour of non-Beltrami fluids guaranteed by
the Structure Theorem, the flowlines of Beltrami fluids can be
arbitrarily knotted \cite{dombre} \cite{etnaghr1} \cite{etnaghr2}
\cite{etnaghr3}. Generically a fluid will be non-Beltrami, but
Beltrami fluids do arise, for example, as ``force-free'' fields in
magneto-hydrodynamics \cite{chandra}. 

In the relativistic setting, stationary perfect fluids have long been
studied and while much progress has been made, a convincing overview has
yet to emerge. The literature in this area is rather
extensive and we mention just some topics pertinent to our result -
for a review see \cite{chinagr2}. In
particular, there has been much progress in finding and investigating
local solutions that satisfy various simplifying assumptions. These can
range from the weak (like energy conditions) to the strong (like
cylindrical symmetry), or a combination of both. 

Without doubt the
most commonly imposed conditions are those of axial symmetry and
circular flow, usually supplemented by some other restriction
\cite{kramer} \cite{kras1} \cite{kras2} \cite{wahl} \cite{chinagr1}.
This setting comes closest to Newtonian theory and thus we are
entitled to attempt to physically interpret the geometric quantities
that arise.  As has been shown, however, even in this case, care must
be taken \cite{chinapareja}. A further down side is that it is not
clear which of the properties these local solutions possess are just
products of their specialty. Thus, failure to match globally to the
vacuum Kerr solution, a difficulty that every known stationary perfect
fluid possesses, may simply be an artifact of their axial symmetry or
circularity. 
 
Aside from exact solutions, there has been progress on general matching
conditions with the Kerr solution \cite{roos1} \cite{roos2}, as well as
results on {\it a priori} estimates on the equatorial radius
\cite{klenk}, positivity of angular momentum density \cite{hanawini}
and asymptotics \cite{caporali}. In all cases axial symmetry is
assumed, or the fluid is assumed to have zero pressure, i.e. the body
consists of dust.  

It is within this context that our Structure Theorem above should be
viewed. For here, the assumptions made on the fluid are nothing
more than one would impose on any physically reasonable model for an
isolated stationary fluid body. Moreover, such a stationary perfect
fluid is precisely what is required to model the equilibrium states of
rotating fluid bodies \cite{ehlers} \cite{lind}. 

Examples of both types of cells in the Theorem have appeared in the
literature. The remarkable perfect fluid source for the NUT metric
discovered by Luk\'{a}cs \cite{lukacs} has flowlines given by the Hopf
fibering of $S^3$, which lie on invariant tori. The second type of
cells arise, for example, when one considers cylindrically symmetric flows
\cite{davidson1} \cite{skla} \cite{wini} \cite{vishawini}. 
The decomposition of the full Einstein field
equations parallel and transverse to the foliation is worthy of
careful consideration.

It is hoped that the Structure Theorem can yield a new approach to
finding a source for the Kerr metric.  It indicates
that the simplest such source must be a solid torus and the fact
that the singularity in the maximally extended Kerr solution is ring
shaped supports such a conclusion.

Another point worthy of further consideration is the Beltrami
condition. In the non-relativistic case, Beltrami
flows exhibit chaotic dynamics and are closely related to contact
geometry. The natural definition in the relativistic case afforded us
by the Structure Theorem is a good deal more complex and its
relationship with contact structures is unclear. It nevertheless
raises the question as to whether relativistic effects ``tame'' 
Beltrami flows, or whether chaotic regimes persist.

\end{document}